\documentclass[12pt,reqno,addw]{article}
\usepackage[cp1251]{inputenc}
\usepackage[russian]{babel}

\begin{document}


\noindent УДК 62-505.3; 530.145.   PACS  03.65 - w; 02.30.Yy.

\vspace*{5mm}

\begin{center}
{\bf УПРАВЛЕНИЕ КВАНТОВЫМИ СИСТЕМАМИ И ТЕОРИЯ ОПТИМАЛЬНОГО
УПРАВЛЕНИЯ}
\vspace*{5mm}

{\bf В.Ф.~Кротов}
\vspace*{5mm}

{Институт проблем управления  РАН, Москва, Профсоюзная, 65}\\
{Тел.: (495) 334-91-59, Е-mail: vfkrotov@ipu.ru}
\end{center}
\vspace*{5mm}

Проблема управления квантовыми системами (КС)~--- одна из наиболее
актуальных   научно-технических проблем, связанная с новейшими
физическими нанотехнологиями. Конечная цель здесь~--- создание
регулярных методов синтеза переменного электрического поля
(лазерного излучения), управляющего микроскопическими состояниями
атомов и молекул. Успешное внедрение на уровень квантовых систем
современных техник управления, инструментария лазерной физики и
квантовой электроники, реализующего ``хирургию'' атомов и молекул,
открыло уникальные возможности для создания новых материалов,
миниатюризации компьютерной памяти, и других технологий. Эта
проблема оказалась также благодарным объектом для математиков,
специализирующихся в области теории управления. Теория методов
синтеза управления квантовыми системами основана на некоторых
идеях теории оптимального управления, адекватных свойствам этих
систем. Здесь эти идеи анализируются и развиваются применительно к
особенностям КС. Помимо сложности фазового пространства,
выражающегося в высокой размерности аппроксимирующих схем, это
характерная нелинейность задачи оптимального управления для таких
систем и следующие из  нее сингулярности решений, создающие
известные пробелы в возможностях оптимизации. На основе полученных
результатов эти пробелы устраняются, расширяется сфера применения
алгоритмов синтеза управления, повышается их эффективность и
быстродействие.

\section*{1. Постановка задачи}

\hspace{0.5cm} В квантовой механике (КМ) синтезированы
детерминированная динамика волновой функции (ВФ)    и
статистическая связь последней с наблюдаемыми величинами:
динамическая и статистическая части КС. Синтез
управления целиком связан с первой, вторая же обусловливает выбор
критериев. Пусть $x$~---  вектор координат  системы, пробегающий
область  $\Omega$;   $\Psi =\Psi(x)$~--- ее ВФ (комплексная),
$\Psi(x)=0$  на границе области $\Omega$; ВФ рассматривается как
элемент комплексного гильбертова пространства $H$ с нормой
$L_2(\Omega)$  и соответствующим произведением
\begin{equation}\label{1a1}
(\Psi_1,\Psi_2)=\int \Psi^*_1(t,x)\Psi_2(t,x)d\Omega,
\end{equation}
где $^{*}$~--- значок комплексного сопряжения,   $d\Omega$~---
элементарный объем координатного пространства. Полное описание
динамического состояния КС  содержится в ВФ  $\Psi(x)$, но здесь в
этом качестве удобнее использовать пару  $\Psi(x)$, $\Psi^*(x)$.
Динамика КС описывается дифференциальными уравнениями Шрёдингера.
\begin{equation}\label{2a1}
\begin{array}{c}
d\Psi(t)/dt=-i/\hbar H^{\wedge}(u)\Psi(t),\quad
d\Psi^*(t)/dt=i/\hbar H^{\wedge}\Psi^*(t),\\[2mm]
\Psi(0,x)=\xi(x),
\end{array}
\end{equation}
где    $t\in [0,T]$~--- время;   $\hbar$~--- постоянная Планка;
$\xi (x)$~--- заданная функция; $u(t)$~--- электрическое поле,
зависящее только от времени, свободно выбираемое в пределах $a\leq
u(t)\leq b$; $H^{\wedge}(u)=H^{\wedge 0}+H^{\wedge 1}u$~---
гамильтонов оператор системы; например, для частицы массы $m$,
управляемой монохроматической радиацией с переменной амплитудой:
$H^{\wedge}(u)=\linebreak =K^{\wedge}+V^{\wedge}-M^{\wedge}u$,
$K^{\wedge}=(\hbar /2m)\triangle$~--- оператор (дифференциальный)
кинетической энергии, $V^{\wedge}=V(x)$~--- потенциальная энергия,
$M^{\wedge}$~--- оператор дипольного момента. Все операторы в
КМ~--- эрмитовы, а функционалы~--- квадратичные формы (КФ). При
любом выборе управления $u(t)$ система (\ref{2a1}) имеет
динамический инвариант
\begin{equation}\label{3a1}
\| \Psi (t)\|^2=(\Psi (t),\Psi (t))={\rm const}=1.
\end{equation}
Последнее равенство диктуется статистическим смыслом ВФ. Т.\,е.
область достижимости управляемой КС есть единичная сфера в $H$.

Необходимо выбрать функцию $u(t)$  так, чтобы она минимизировала
заданный функционал:
\begin{equation}\label{4a1}
I[u(t)]=-(\Psi (T),L\Psi (T))\to \min ,
\end{equation}
адекватный  целям управления. Здесь   $L$~--- соответствующий
оператор; знак минус обусловлен тем, что во многих случаях
максимизируется положительная квадратичная форма. Это~--- задача
оптимального управления для процессов в функциональном
пространстве состояний $H$.

{\bf Замечание}. Иногда применяются и критерии более общего вида:
\begin{equation}\label{5a1}
J[u(t)]=I+\int\limits^T_0 \left\{\left(\Psi^*(t)N[u(t)]\Psi
(t)\right)dt+f[u(t)]\right\}d\Omega \to \min ,
\end{equation}
где   $N$~--- эрмитов оператор, зависящий от  $u$,   $f[u]$~---
заданная функция, функционал $I$  определен (\ref{4a1}). Особенно
часто:
\begin{equation}\label{6a1}
J=I+I_{\beta}[u(t)];\ I_{\beta}[u(t)]=\int\limits^T_0 \beta
u^2(t)dt,\quad \beta >0
\end{equation}
(см. об этом ниже).

\section*{2. Особенности задачи}

Рассмотрим ординарную задачу оптимального управления
\begin{equation}\label{7a1}
I(v)=\int\limits^T_0
f^0\left(t,x(t),u(t)\right)dt+F\left(x(T)\right)\to
\min\limits_{v\in D}
\end{equation}
\begin{equation}\label{8a1}
dx/dt=f(t,x,u);\quad x(0)=\xi ;\quad x\in E^n;\quad u\in U,
\end{equation}
где   $E^n$~--- эвклидово пространство размерности $n$  с
произведением $xy=\linebreak ={\displaystyle\sum}^n_1x^iy^i$;
вектор $\xi$, вектор-функция $f(t,x,u)$, функции $F(x)$,
$f^0(t,x,u)$ и компактное множество $U$ заданы,   $D$~---
множество допустимых процессов $v=\left(x(t),u(t)\right)$,
удовлетворяющих (\ref{8a1}). В соответствии со сказанным в п.~1
ограничимся случаем:   $u$~--- скаляр, $u\in [a,b]$,
$f^0(t,x,u)=\beta u^2$, $\beta \geq 0$. Причем случаи $\beta =0$ и
$\beta >0$, как мы увидим, алгоритмически существенно различны.

Для практических вычислений гильбертов вектор $\Psi$
аппроксимируется вектором  $\Psi'\in E^n$, и задача (\ref{2a1}),
(\ref{4a1})~--- задачей (\ref{7a1}), (\ref{8a1}). Адекватная
размерность вектора $\Psi'$ очень велика: $n\approx 10^4-10^6$ для
молекулы с 6~--~7 степенями свободы. Это~--- первая особенность
задачи, фильтрующая выбор алгоритмов. Вторая,~--- комплексность
ВФ,~--- не столь существенна для алгоритмизации; третья~---
отсутствие ограничений на состояние, включая терминальные
ограничения; четвертая: (\ref{2a1}) суть линейные однородные
уравнения с управляемыми коэффициентами специального вида, в
представлении (\ref{8a1}):
\begin{equation}\label{9a1}
{\rm a)}\ f(t,x,u)=C(t,u)x,\quad {\rm b)}\ C=A(t)+B(t)u,
\end{equation}
где  $A$, $B$~--- матричные операторы,  $E^n\to E^n$, зависящие от
$t$.

Для различных модификаций задачи (\ref{2a1}), (\ref{4a1}),
(\ref{6a1}) выведены необходимые условия оптимальности типа
краевых задач для уравнений Эйлера и принципа максимума Понтрягина
\cite{bib1} (текст и ссылки).  Но непосредственное их
использование для синтеза управления возможно только применительно
к ``игрушечным'' моделям малой размерности. Эффективный инструмент
для решения задач (\ref{7a1}), (\ref{8a1}) с большим  $n$, и
соответственно, для (\ref{2a1}), (\ref{4a1}): методы итеративного
улучшения программы управления. В настоящее время применительно к
КС используются два таких алгоритма: градиентный (ГрМ) и
глобальный (ГлМ). Перечисленные особенности задачи создают удобные
предпосылки для их применения.

\section*{3. Методы последовательного улучшения программы
управления}

Выделим из (\ref{7a1}) ``подзадачу улучшения''. Имеется
допустимый, но не оптимальный процесс
$v_0=\left(x_0(t),u_0(t)\right)\in D$. Требуется улучшить его,
отыскав процесс $v=\left(x(t),u(t)\right)\in D$, такой, что
$I(v)<I(v_0)$.

Повторяя эту операцию, получим улучшающую последовательность
$\{v_s\}\subset D$, $I(v_{s+1})<I(v_s)$. Предел $\lim\limits_{s\to
\infty}I(v_s)=\inf\limits_{v\in D}I(v)$ мы здесь не исследуем,
сосредоточившись на операции улучшения и некоторых оценках
последовательности.

Унифицированное описание интересующей нас группы методов дает
техника достаточных условий оптимальности \cite{bib2}. Запишем
семейство представлений функционала $I(v)$  с непрерывной, дважды
дифференцируемой по $x$ функцией $\varphi (t,x)$ в качестве
параметра:
\begin{equation}\label{10a1}
I(v,\varphi )=G\left(x(T)\right)-\int\limits^T_0
R\left(t,x(t),u(t)\right) dt\to \min\limits_{v\in D}
\end{equation}
\begin{equation}\label{11a1}
R(t,x,u)=H(t,\varphi_x,x,u)+\varphi_t;\ H(t,p,x,u)=pf(t,x,u)-f^0(t,x,u)
\end{equation}
\begin{equation}\label{12a1}
G(x)=F(x)+\varphi (T,x)-\varphi (0,\xi).
\end{equation}
Здесь и далее нижний значок означает дифференцирование по
соответствующей переменной.

Существенную роль в решении задачи улучшения играют сопряженные
уравнения
$$
R_x[t,x_0(t),u_0(t)]=dy(t)/dt+yf\left(t,x_0(t),u_0(t)\right) =0;
$$
\begin{equation}\label{13a1}
G_x\left(x_0(T)\right)=y(T)+F_x\left(x_0(T)\right)=0,
\end{equation}
которые определяют вектор-функцию
$y(t)=\varphi_x\left(t,x_0(t)\right)$.
\medskip

{\bf 3.1. Градиентный метод}. Пусть $\varphi (t,x)=y(t)x$, $y(t)$
удовлетворяет (\ref{13a1}). Будем искать процесс $v$, достаточно
близкий к   $v_0$, чтобы знак разности $\triangle I=I(v)-I(v_0)$
был тот же, что и у ее главной линейной части
\begin{equation}\label{14a1}
\delta I(v)=-\int\limits^T_{\tau} R_u\left(t,x_0(t),u_0(t)\right)
\delta u_0(t)dt,
\end{equation}
где   $R_u\left(t,x_0(t),u_0(t)\right)
=H_u\left(t,y(t),x_0(t),u_0(t)\right)$, $\delta u_0(t)$~---
вариация программы управления. Зададим $\delta u_0(t)$ так, что
правая часть (\ref{14a1})  положительна. Пусть при достаточно
малых $\varepsilon >0$: $u(t,\varepsilon )=u_0+\varepsilon \delta
u_0\in U$, $\forall \ t$, и траектория $x(t,\varepsilon )$
определена управлением $u(t,\varepsilon )$ в силу (\ref{8a1}).
Тогда существует $\varepsilon >0$, такое что $I(v)<I(v_0)$,
$v=v(\varepsilon)$.

Таким образом, процесс улучшения программы управления $u_0(t)$
сводится к следующим шагам: 0. инициализация: дано $u_0(t)$,
находим $x_0(t)$, решая задачу Коши (\ref{8a1})  с   $u=u_0(t)$;
1. находим $y(t)$ и  $R_u\left(t,x_0(t),u_0(t)\right)$, решая
линейную задачу Коши (\ref{13a1}), либо, не запоминая $x_0(t)$,
$t<T$,~--- задачу Коши для уравнений (\ref{8a1}), (\ref{13a1}) при
$x(t)=x_0(t)$ с начальными условиями $x(T)=x_0(T)$ и (\ref{13a1});
2. установим вариацию программы управления $\delta u_0(t)$ так,
чтобы правая часть (\ref{14a1}) была положительной; 3. Для
различных $\varepsilon>0$, найдем решения $x(t,\varepsilon)$
задачи Коши (\ref{8a1}) при   $u=u_0+\varepsilon \delta u_0$.
Величину $\varepsilon$ следует выбрать так, чтобы выполнялось
$I(v)<I(v_0)$, $v=v(\varepsilon )$.

Выражение (\ref{13a1}) дает градиент функционала $I(u)$ в
пространстве уп\-равлений $u(t)$. Эти методы развивались в работах
Келли и др. \cite{bib3}, Энеева \cite{bib4}, Брайсона \cite{bib5},
и других.
\medskip

{\bf 3.2. Глобальный метод улучшения управления}. Обозначим
\begin{equation}\label{15a1}
\widetilde{u}(t,x)=\mathop{\arg\max}\limits_{u\in
U}R(t,x,u)=\mathop{\arg\max}\limits_{u\in U}H(t,\phi_x,x,u).
\end{equation}
Опишем операцию улучшения. Инициализация подобна п.~0. из~3.1.
\begin{enumerate}
    \item Конструируем функцию  $\varphi (t,x)$, такую что
\begin{equation}\label{16a1}
R\left(t,x_0(t),u_0(t)\right) =\min_x
R\left(t,x,u_0(t)\right),\quad \forall \ t\ne T
\end{equation}
\begin{equation}\label{17a1}
G\left(x_0(T)\right)=\max_x G(x).
\end{equation}
\item Из (\ref{15a1}) находим $\widetilde{u}(t,x)$  и определяем
процесс $v=\left(x(t),u(t)\right) =\linebreak
=\widetilde{u}\left(t,x(t)\right)\in D$ согласно уравнению и
начальным условиям (\ref{8a1}).
\end{enumerate}

{\bf Теорема}, \cite{bib6}. {\em Имеем}\ : $I(v)\leq I(v_0)$. {\em
Если условие} $R\left(t,x_0(t),u_0(t)\right)=\linebreak
=\max\limits_x R\left(t,x_0(t)<u\right)$, $\forall \ t\ne T$, {\em
не выполняется}, {\em то} $I(v)<I(v_0)$.
\medskip

Воспроизводя эту операцию, получим улучшающую последовательность
$\{v\} \subset D$. Основное звено этого метода, ~---
конкретный способ построения функции $\varphi (t,x)$ на каждой
итерации. Их варианты с подробным описанием см.~в  \cite{bib7},
\cite{bib8}, \cite{bib2}.
\medskip

{\bf 3.3. Линейные системы с управляемыми коэффициентами}. Пусть
справедливо (\ref{9a1}) a). Зададим: $\varphi (t,x)=y(t)x$. Имеем
\begin{equation}\label{18a1}
\begin{array}{c}
R(t,x,u)=\left[dy(t)/dt+C^T(t,u)y\right]x-\beta u^2;\\[2mm]
G(x)=F(x)+y(T)x-y(0)x_0.
\end{array}
\end{equation}
\begin{equation}\label{19a1}
\begin{array}{c}
R_x\left[\left(t,x_0(t),u_0(t)\right)\right]=dy/dt+C^T\left(t,u_0(t)
\right)y=0;\\[2mm]
y(T)+F_x\left(x_0(T)\right)=0.
\end{array}
\end{equation}
Здесь $(\cdot )^T$~--- значок  транспонирования. Функция
$R(t,x,u)$ выпукла по $x$.Таким образом, первое условие
(\ref{19a1}) необходимо и достаточно для (\ref{16a1}). Если
функция $F(x)$  вогнута, то второе условие (\ref{19a1}) необходимо
и достаточно для (\ref{17a1}). Т.\,е. функция $\varphi
(t,x)=y(t)x$ удовлетворяет (\ref{16a1}), (\ref{17a1}) и глобальный
улучшающий алгоритм воспроизводит алгоритм градиентного метода с
заменой  $u=u_0+\varepsilon \delta u_0$ в п.п.~2, 3 на
$u=\widetilde{u}(t,x)=\linebreak =\mathop{\arg\max}\limits_{u\in
U}\left[xC^T(t,u)y-\beta u^2\right]$, при этом существенно упрощая
последний, поскольку исключается настроечный параметр
$\varepsilon$, и бесконечно малые теоретически шаги заменяются
конечными.
\medskip

{\bf 3.4. Особые режимы}. Пусть справедливо (\ref{9a1}) b) и
$\beta =0$. Имеем:
\begin{equation}\label{20a1}
\begin{array}{c}
dx/dt=(A+Bu)x;\quad dy/dt=-(A^T+B^Tu)y;\\[2mm]
H(t,y,x,u)=yAx+K(t,y,x)u;
\end{array}
\end{equation}
\begin{equation}\label{21a1}
\begin{array}{c}
K=yBx;\quad \widetilde{u}(t,x)=a,\quad K<0;\\[2mm]
\widetilde{u}(t,x)=b,\quad K>0;\quad \widetilde{u}(t,x)={\rm
var},\quad K(t,x)=0.
\end{array}
\end{equation}
Пусть при выполнении п.~3 операции улучшения при $t=t_1:K(t_1)=0$.
Значение $u(t_1)$ оказывается не фиксировано. Доопределим его
значением $u(t)=u_{\rm sing}(t)$, реализующим равенство
$K\left(t,x(t)\right)=0$, $t\geq t_1$. Дифференцируя последнее,
получим с учетом (\ref{20a1}), (\ref{21a1}):
\begin{equation}\label{22a1}
u_{\rm sing}(t)=u_0(t)+\left[(BA-AB)^Ty(t)\right]x(t)/y(t)B^2x(t).
\end{equation}
Это решение назовем особым, или сингулярным, режимом улучшенного
управления. Оно появляется в улучшающей последовательности в
качестве полуфабриката особого режима оптимального управления,
если последний содержится в оптимальном процессе, и ограничено
условием  $u_{\rm sing}\in [a,b]$. Этот режим получен в рамках
глобального метода. Градиентный метод здесь неприменим:
$R_u\left(t,x_0(t),u_0(t)\right)=K\left(t,x_0(t)\right)=0$,
соответственно, улучшающий сдвиг управления $\delta u$ не
определен, а управление $u_{\rm sing}(t)$  не следует из логики
метода и, вообще, противоречит его условиями (малость $\delta u$).

Характерна возможная неединственность улучшенного процесса
$v=\linebreak =x(t),\ u(t)$. При определенных условиях имеется
альтернатива: ``сойти'' с особого режима или остаться на нем.
Именно, если в момент $t\geq t_1$ выполняется одно из неравенств
\begin{equation}\label{23a1}
f(t,x,a)K_x(t,x)<0;\quad f(t,x,b)K_x(t,x)>0,
\end{equation}
то соответственно
\begin{equation}\label{24a1}
u(t+0)=\{u_{\rm sing}(t)\ \mbox{или}\ a\} ;\quad u(t+0)=\{u_{\rm
sing}(t)\ \mbox{или}\ b\} .
\end{equation}

{\bf Замечание 1}. Терминальные условия (\ref{19a1}) и
$K\left(T,y(T)x_0(T)\right)=0$ совместны только при специальных
значениях $x_0(T)$. Поэтому, вообще, для выполнения 2-го и
последующих улучшений необходимо, чтобы участок улучшаемой
траектории, примыкающий к $t=T$, был неособым:\linebreak
$K\left(T,y(T),x(T)\right)\ne 0$.

{\bf Замечание 2}. Пусть управление имеет несколько компонент:
\begin{equation}\label{25a1}
u=(v,w),\quad w\in [a,b]\subset R^1;\quad C(t,u)=A(t,v)+B(t,v)w.
\end{equation}
Изложенный алгоритм применим к улучшению по  $w$  при
фиксированном $v(t)$, так что можно считать:  $A=A(t)$,  $B=B(t)$.
В последовательности эти шаги чередуются с улучшениями других
компонент управления.

\section*{4. Квантовые системы}

Постановка задачи ~--- в п.~1. Она относится к классу задач п.п.~
2.3, 2.4  в соответственно обобщенном пространстве состояний. Роль
матричных операторов  $A$, $B$  играют операторы  $H^{\wedge 0}$,
$H^{\wedge 1}$, действующие в $H$. Их дополнительное свойство~---
самосопряженность, и как следствие,~--- (\ref{3a1}).
\medskip

{\bf 4.1. Условия применимости методов улучшения}. ГрМ применим к
задаче (\ref{4a1}), только на ранних итерациях и если поиск
начинается далеко от экстремума и не сказываются эффекты особого
режима, либо~--- при отсутствии последнего в составе оптимального
процесса (что не характерно). При наличии добавки (\ref{6a1}) он
применим без оговорок (см. об этом ниже).

Применимость глобального улучшения регламентируется в п.~2.2
требованием вогнутости функции $F(x)$, чему соответствует
неотрицательность формы  $\left(\Psi (T),L\Psi (T)\right)$.
Покажем, что для КС оно смягчается до требования
знакоопределенности последней и, как правило, выполняется.
Наиболее характерные критерии оптимальности~--- максимум или
минимум  вероятности  $P[\Psi (T),Q]$ того, что к моменту $T$
значения наблюдаемых величин, или соответствующие состояния,
окажутся в пределах заданного множества $Q$. Пусть наблюдаемые~---
энергия или импульс, имеющие собственные значения $\lambda_i$  и
функции $\Psi_i$. Имеем: $P[\Psi (T),Q]=\linebreak =\sum_i|
(\Psi_i,\Psi (T))|^2$, $i:\lambda_i\in Q$. Это~--- положительная
КФ. Задача $I[u(t)]=\linebreak =-P[\Psi (T),Q]\to \min$,
соответствующая максимизации $P$, отвечает требованиям
применимости метода. Задача $P[\Psi (T),Q]\to \min$ не отвечает
им, но ей эквивалентна задача $I[u(t)]=-P[\Psi (T),Q']\to \min$,
где $Q'$~--- дополнение $Q$ до числовой оси. Последняя обладает
необходимой выпуклостью. Аналогично обстоит дело, если
наблюдаемые~--- координаты, и соответственно $P[\Psi
(T),Q]=\int_Q\Psi^*(T,x)\Psi (T,x)d\Omega$.Таким образом, и
максимизация, и минимизация $P[\Psi (T),Q]$ реализуема этим
методом либо непосредственно, либо альтернативной заменой  $Q\to
Q'$.
\medskip

{\bf 4.2. Операция улучшения}. Запишем базисные конструкции этих
методов, учитывая особенности задачи. Улучшающая функция $\varphi
(t,\Psi ,\Psi^*)$ теперь~--- линейный действительный функционал:
$\varphi =\left(\chi (t),\Psi\right)+\linebreak
+\left(\chi^*(t),\Psi^*\right)={\rm Re}\left(\chi
(t),\Psi\right)$, где $\chi (t)$~--- заданная вектор-функция,
$[0,T]\to \linebreak \to H$, как и соответствующие ей в силу
(\ref{13a1}):
$$
\begin{array}{c}
G(\Psi ,\Psi^*)=-(\Psi ,L\Psi )+\left(\chi (T),\Psi
\right)+\left(\chi^*(T),\Psi^*\right) \\[2mm]
R(t,\Psi ,\Psi^*,u)=K^0(t,\Psi ,\Psi^*)+K^1(t,\Psi ,\Psi^*)u-\beta
u^2+({\dot \chi},\Psi)+({\dot \chi}^*,\Psi^*),\\[2mm]
K^j(t,\Psi ,\Psi^*)=i/\hbar
\left[\left(\chi^*(t),H^j\Psi^*\right)-\left(\chi
(t),H^j\Psi\right)\right],\ j=0,1.
\end{array}
$$
Дифференциальные уравнения итераций:
\begin{equation}\label{26a1}
{\rm a)}\ {\dot \chi}(t)-(i/\hbar )u_0(t)H^{\wedge}\chi
(t)=0;\quad {\rm b)}\ {\dot \Psi}(t)=-(i/\hbar
)\widetilde{u}(t,\Psi ,\Psi^*)H^{\wedge}\Psi (t),
\end{equation}
\begin{equation}\label{27a1}
\begin{array}{c}
\beta =0:\widetilde{u}(t,\Psi ,\Psi^*)=\left\{%
\begin{array}{ll}
    a, & K^1<0 \\
    b, & K^1>0 \\
    u_{\rm sing}, & K^1=0 \\
\end{array}
\right.;\\[7mm]
\beta >0:\widetilde{u}(t,\Psi ,\Psi^*)=\left\{%
\begin{array}{ll}
    a, & u'<a \\
    u', & a<u'<b \\
    a, & u'>b \\
\end{array}%
\right.,
\end{array}
\end{equation}
$$
\begin{array}{c}
u_{\rm sing}=u_0(t)-\left[\left((H^1H^0-H^0H^1)\chi (t),\Psi\right)
\right]+\\[2mm]
+\left[\left((H^1H^0-H^0H^1)\chi^*,\Psi^*\right)
\right]/\left[((K^1)^2\chi ,\Psi )+((K^1)^2\chi ,\Psi )\right] ;\\[2mm]
u'=-K^1/2\beta ;\quad R_u(t,\Psi ,\Psi^*,u)=K^1(t,\Psi
,\Psi^*)-2\beta u.
\end{array}
$$

В соответствии с п.п.~2.3, 2.4, операция глобального улучшения
будет следующей. 0. Инициализация, прямая прогонка.  Выбирается
функция $u=u^{(0)}(t)$   и интегрируется уравнение Шредингера
(\ref{2a1}) для $\Psi$ c начальным значением  $\xi (x)$,
получается траектория $\Psi^0(t)$  и соответствующее значение
функционала $I^{(0)}$.

1. Обратная прогонка. Решается задача Коши (\ref{26a1}) a), $\chi
(T)\!=\!-L\Psi_0(T)$, и находится $\chi^0(t)$.

2. Прямая прогонка. Решаем задачу Коши (\ref{26a1}),  $\Psi
(0,x)=\xi (x)$, $\chi (0)=\linebreak =\chi^0(0)$, воспроизводя
$\chi^0(t)$ и определяя новую траекторию $\Psi (t)$, управление
$u(t)=\widetilde{u}(t,\Psi (t))$ согласно (\ref{27a1}), и
соответствующее значение функционала $I$  или $J$.

Применение операции градиентного улучшения к задаче  (\ref{6a1})
воспроизводит хорошо известные схемы. Методы гарантируют только
нахождение локального минимума (понтрягинская экстремаль).
\medskip

{\bf 4.3. Энергия поля как регламентирующий фактор управления}.
Функционал $I_{\beta}$ интерпретируется как расход энергии на
управление, а минимизация суммы    $J$~---  это компромисс между
качеством достижения главной цели (\ref{4a1}) и экономией энергии.
Но содержание этого компромисса требует прояснения. Если энергия
лазерного излучения есть регламентирующий фактор управления, то
его следует ввести в постановку задачи в виде ограничения:
\begin{equation}\label{28a1}
dz/dt=u^2(t);\quad z(0)=0;\quad z(T)\leq a,
\end{equation}
где $a$~--- предельное допустимое значение энергии. Этому
ограничению отвечает сопряженная функция:  $\beta (t)={\rm
const}$; $\beta =0$, $z(T)<a$; $\beta >0$, $z(T)=a$.

Решается задача улучшения при $\beta =0$  и проверяется
(\ref{28a1}). Если $z(T)\leq a$, то ограничение (\ref{28a1})
несущественно; если  $z(T)>0$, то решается семейство задач
улучшения с параметром  $\beta >0$, который подбирается так, что
$z(T)=a$. Как видим, редукция $I\to J(\beta )$ не тождественна
учету ограниченности энергии.
\medskip

{\bf 4.4. Сравнение методов глобального и градиентного
последовательного улучшения}. ГрМ имеет локальный характер, улучшение требует семейства прогонок
с настроечным параметром $\varepsilon$ и гарантируется только при
малых его значениях. Соответственно, сходимость происходит
медленно. ГлМ не содержит настроечных коэффициентов и реализуется
единственной парой прогонок: обратной и прямой.  ГрМ не содержит
достаточных инструментов для улучшения и генерирования особого
режима, т.\,е. ~--- оптимизации КС с критерием качества
(\ref{4a1}): оптимальная траектория состоит из кусков $u=a$, $u=b$
и особого режима $K^1(t)\equiv 0$, для которого градиент и,
соответственно, улучшающий сдвиг управления $\delta u$  не
определен, а управление $u_{\rm sing}(t)$ не следует из логики
метода и, вообще, не допускается его условиями (малость  $\delta
u$). Поэтому при пользовании ГрМ обычно к функционалу (\ref{4a1})
добавляется слагаемое (\ref{26a1}).  Оно регуляризует ГрМ, но
вносит неясность в физический смысл критерия (см. выше) и
обусловливает недоиспользование ресурса управления для главной
цели (\ref{4a1}): ухудшается и сходимость последовательности
улучшений сравнительно с ГлМ, и качество конечного результата. ГлМ
не имеет этих осложнений и применяется либо напрямую к исходной
задаче (\ref{4a1}), либо к редуцированной, хотя и в нем  не все
обстояло благополучно с особым режимом. Теоретически сингулярная
дуга улучшенной траектории $K(t)=0$  автоматически реализуется ГлМ
как скользящий режим. Поэтому в \cite{bib2} предлагалось не
выделять ее специально. Однако реализация такого подхода
оказывается связана с существенными вычислительными трудностями.
Регулярное управление $u_{\rm sing}(t)$, полученное здесь, снимает
эти трудности.

ГрМ применялся к КС еще до появления ГлМ, лучше освоен физиками, и
большинство задач решается им. Применение ГлМ к проблемам
управления КС было предложено в \cite{bib9}, и к настоящему
времени его можно считать достаточно внедренным. Его преимущества
нашли подтверждение в исследованиях управления квантовым
состоянием молекул и задач физической химии, магнитного резонанса,
задач квантовой оптики, и других актуальных направлений,
опубликованных в \cite{bib10}~--~\cite{bib15}  и других работах.
Следует выделить \cite{bib13}, где этим методом в модификации
\cite{bib8} оптимизируется нелинейная КС, выпадающая за рамки
уравнений (\ref{2a1}) (бозонный конденсат). Интересный
сравнительный анализ был сделан в \cite{bib10, bib11}. Показано,
что ГлМ существенно уменьшает необходимое количество вычислений,
если поиск начинается далеко от экстремума.  Численные
эксперименты на нескольких КС с критерием (\ref{6a1}) хорошо это
демонстрируют. Сходимость происходит очень быстро в самом начале
поиска.

Автор благодарит А.В.~Булатова и А.В.~Горшкова за консультации и
обсуждение работы. Аббревиатуры: КМ~--- квантовая механика; ВФ~---
волновая функция; КС~--- квантовая система; (ГлМ)~--- глобальный
метод; (ГрМ)~--- градиентный метод.

\end{document}